\documentclass{ws-ijmpa}
    \newcommand{\ba}{\begin{eqnarray}}
    \newcommand{\ea}{\end{eqnarray}}
    \newcommand{\be}{\begin{equation}}
    \newcommand{\ee}{\end{equation}}

    \newcommand{\AmS}{{\protect\the\textfont2%
  A\kern-.1667em\lower.5ex\hbox{M}\kern-.125emS}}

\newcommand{\bn}{{\bf n}}

\newcommand {\bk} {{\mathbf k}}
\newcommand {\bfr} {{\mathbf r}}

\begin{document}

\markboth{Chuan Liu, Xu Feng, Song He} {Two Particle States and
the $S$-matrix Elements in Multi-channel Scattering}

\catchline{}{}{}{}{}
%
%%%%%%%%%%%%%%%%%%%%%%%%%%%%%%%%%%%%%%%%%%%%%%%%%%%%%%%%%%%%%%%%%%%%

\title{TWO PARTICLE STATES IN A BOX AND THE \\
$S$-MATRIX IN MULTI-CHANNEL SCATTERING
%\footnote{For the title, try not to use more than
%3 lines. Typeset the title in 10 pt roman, uppercase and
%boldface.}
}

\author{\footnotesize CHUAN LIU, XU FENG and SONG HE
%\footnote{ Typeset names in 8 pt roman, uppercase. Use the
%footnote to indicate the present or permanent address of the
%author.}
}

\address{School of Physics, Peking University\\
Beijing 100871, China
%\footnote{State completely without
%abbreviations, the affiliation and mailing address, including
%country. Typeset in 8 pt italic.}
}

\maketitle

%\pub{Received (Day Month Year)}{Revised (Day Month Year)}

\begin{abstract}
 Using a quantum mechanical model,
 the exact energy eigenstates for two-particle two-channel
 scattering are studied in a cubic box
 with periodic boundary conditions.
 A relation between the exact energy eigenvalue in
 the box and the two-channel $S$-matrix elements in the continuum
 is obtained. This result can be viewed as a
 generalization of the well-known L\"uscher's formula which
 establishes a similar relation in elastic scattering.

\end{abstract}

\section{Introduction}    %) A SECTION HEADING

 Scattering experiments play an important role
 in the study of interactions among particles.
 In the case of low-energy hadron-hadron scattering,
 experimental results on the phase shifts are available~\cite{E86500:pipi}.
 On the theoretical side,
 low-energy hadron-hadron
 scattering can be studied with a non-perturbative
 method, like Lattice QCD.
 In a typical lattice calculation, energy eigenvalues of two-particle states
 with definite symmetry are obtained. Therefore, it is important to
 relate the energy eigenvalues which are available through
 lattice calculations to the scattering phases which are obtained
 in the scattering experiments.  This was accomplished
 in a series of papers by L\"uscher
 \cite{luscher91:finitea}
 for a cubic box topology. This formula, now known as
 L\"uscher's formula, has been utilized in a number of
 applications, e.g. linear sigma model in the broken phase \cite{Zimmermann94},
 and also in quenched~\cite{gupta93:scat} and
 unquenched~\cite{CPPACS03:pipi_phase_unquench} QCD.

 For hadron scattering at low energies, elastic scattering dominates.
 However, when the energy of the scattering
 process exceeds some threshold, inelastic scattering starts
 to contribute. In the case of pion-pion scattering, for example, the scattering
 process is elastic below the four pion and the two kaon
 threshold. Although the four pion threshold is in fact below the
 two kaon threshold, four pion final states will not
 contribute significantly due to its weak chiral coupling to
 the two pion initial state.
 Experimental investigations also supports this argument.
 It is then interesting to study the relation between the
 multi-channel two-particle states
 and the scattering phases in general.
 In this work, we report a relation between the energy
 of a two particle state in a finite cubic box and the scattering
 matrix parameters. Further details are provided in
 Ref.~\cite{chuan05:2channel}.

 \section{The Quantum Mechanical Model of Two Channel Scattering}
 \label{sec:model}

 We study a non-relativistic quantum mechanical model of
 two-channel scattering with the following Hamiltonian:
 \be
 \label{eq:hamiltonian}
 H=\left(\begin{array}{cc}
 -{1\over 2m_1}\nabla^2 & 0 \\
 0 & E_T-{1\over 2m_2}\nabla^2\end{array}\right)
 +\left(\begin{array}{cc}
 V_1(r)      & \Delta(r) \\
 \Delta^*(r) & V_2(r)\end{array}\right)\;.
 \ee
 with the potential vanishes for $r>R$.
 $E_T>0$ designates a positive threshold energy
 of the second channel.

 Energy eigenstates of the Hamiltonian~(\ref{eq:hamiltonian}) with
 energy $E$ can be decomposed into spherical harmonics:
 $\Psi_i(\bfr)=\sum_{l,m} Y_{lm}(\hat{\bfr})
 \psi_{i;lm}(r)$
 where the radial wave-functions $\psi_{i;lm}(r)$ with $i=1,2$ satisfy
 the usual radial Schr\"odinger equation.
 Concerning the coupled radial differential
 equations, the following statement can be proven~\cite{chuan05:2channel}:
 {\em If the matrix valued potential $V(r)$ is
 such that every matrix element of $r^2V(r)$ is
 analytic around $r=0$ and that
 $\lim_{r\rightarrow 0}r^2V(r)=0$, then the
 coupled radial differential equations
 has two finite, linearly independent solutions near $r=0$:
 $u^{(i)}_{l;j}(r)$, with $i=1,2$, $j=1,2$ such that:
 $u^{(i)}_{l;j}(r) \sim r^l\delta_{ij}$.}

 At large $r$ where the potential $V(r)$ vanishes, the
 wave function of the scattering state can be chosen to
 have particular forms:
 \be
 \label{eq:psi1}
 \Psi^{(1)}(\bfr)\stackrel{r\rightarrow\infty}{\longrightarrow}
 \left(\begin{array}{c}
 e^{i\bk_1\cdot\bfr}+
 f_{11}(\hat{\bk}_1\cdot\hat{\bfr}){e^{ik_1r}\over r}\\
 f_{21}(\hat{\bk}_1\cdot\hat{\bfr})\sqrt{{m_2\over m_1}}
 {e^{ik_2r}\over r}\end{array}\right)\;.
 \ee
 This wave function has the property that in the
 remote past, it becomes an incident plane wave
 in the first channel with definite wave vector $\bk_1$.
 It is an eigenstate of the full Hamiltonian with energy: $E=\bk^2_1/(2m_1)$.
 Similarly, if the energy $E>E_T$, one can also
 build another eigenstate of the Hamiltonian $\Psi^{(2)}(\bfr)$, which in past
 becomes an incident wave in the second channel
 with energy $E=E_T+\bk^2_2/(2m_2)$.

 In partial wave analysis, one decomposes the
 coefficients: $f_{ij}$ (scattering applitudes) into spherical harmonics.
 The scattering eigenstate
 in Eq.~(\ref{eq:psi1})
 can aslo be decomposed accordingly with the radial wave-functions:
 \be
 \label{eq:w_def}
 w^{(1)}_l(r) \simeq \left(\begin{array}{c}
 {1\over 2ik_1r}\left[
 S^{(l)}_{11}e^{ik_1r}+(-)^{l+1}e^{-ik_1r}\right] \\
 {1\over 2i\sqrt{k_1k_2}r}\sqrt{{m_2\over m_1}}
 S^{(l)}_{21}e^{ik_2r}
 \end{array}\right)\;,
 \ee
 and a similar equation for $\Psi^{(2)}(\bfr)$ and $w^{(2)}_l(r)$.
 It is obvious that the two radial wave functions
 $w^{(1)}_l(r)$ and $w^{(2)}_l(r)$ thus defined form a
 linearly independent basis. They are hence linear superpositions of the
 general solutions: $u^{(1)}_l(r)$ and $u^{(2)}_l(r)$.
 The converse is of course also true.

 An important physical property is that the
 matrix elements which enter the expansion, namely
 $S^{(l)}_{ij}$, form a $2\times 2$ unitary matrix.
 In practice, this two-channel $S$-matrix is usually
 parameterized as:
 \be
 \label{eq:S_parametrize}
 S^{(l)}(E)=\left(\begin{array}{cc}
 \eta_l e^{2i\delta^l_1} &
 i\sqrt{1-\eta^2_l}e^{i(\delta^l_1+\delta^l_2)} \\
 i\sqrt{1-\eta^2_l}e^{i(\delta^l_1+\delta^l_2)} &
 \eta_l e^{2i\delta^l_2} \end{array}
 \right)\;,
 \ee
 where the parameters: $\delta^l_1$, $\delta^l_2$
 and $\eta_l$ are all real functions of the energy $E$.

 \section{Energy Eigenfunctions on a Torus}
 \label{sec:2channel_finiteV}

 When enclosed in a cubic, periodic
 box with finite extension $L$,
 the Schr\"odinger equation of the system takes a
 similar form except that the
 potential is periodically extended and the
 eigenfunction has to satisfy the periodic boundary condition:
 $[H_0+V_L(\bfr)]\psi(\bfr)=E\psi(\bfr)\;,
 \;\; \psi(\bfr+L\bn)=\psi(\bfr)\;.
 $
 where $H_0$ is the free Hamiltonian
 and the periodically extended potential is:
 $V_L(\bfr)\equiv \sum_\bn V(\bfr+L\bn)$.
 The eigenvalue of the Hamiltonian now
 becomes discrete with smooth eigenfunctions.
 It is also convenient to partition the whole space into two
 regions. In the inner region, every point satisfies the
 condition: $|\bfr|<R$, $mod(L)$. In the outer region:
 $\Omega=\{\bfr|: |\bfr|>R\;,mod(L)\}$,
 where the interaction potential $V_L(\bfr)=0$
 and the Schr\"odinger equation of the system
 reduces to two {\em decoupled} Helmholtz equations:
 $(\nabla^2+k^2_i)\psi_i(\bfr)=0$, $i=1,2$.

 It it easy to see that:
 $\Psi(\bfr;E)=\sum_{lm}
 \left[\sum^2_{i=1} b^{(i)}_{lm}u^{(i)}_l(r)\right]
  Y_{lm}(\bn)$
 solves the Schr\"odinger equation in the
 inner region for $|\bfr|<R$
 with $b^{(j)}_{lm}$ being non-vanishing coefficients.
 In the outer region $\Omega$, the solution
 must be linear superposition of the singular periodic
 solutions of Helmholtz equation~\cite{chuan05:2channel},
 we thus obtain a set of homogeneous linear
 equations. In order to have non-trivial solutions for
 the coefficients, the corresponding matrix has to be singular.
 This condition then gives:
 \be
 \label{eq:main_resultU}
 \left|\begin{array}{cc}
 U^{(1)}_{l'm';lm}-
 S^{(l)}_{11}\delta_{l'l}\delta_{m'm}
 &
 \sqrt{{k_2m_2\over k_1m_1}}S^{(l)}_{21}
 \delta_{l'l}\delta_{m'm}
 \\
 \sqrt{{k_1m_1\over k_2m_2}}S^{(l)}_{12}
 \delta_{l'l}\delta_{m'm}
 &
 U^{(2)}_{l'm';lm}- S^{(l)}_{22}\delta_{l'l}\delta_{m'm}
 \end{array}\right|=0\;,
 \ee
 where the unitary matrices $U^{(i)}$ are specific
 functions of the energy~\cite{chuan05:2channel}.

 In most lattice calculations, the symmetry sector that is
 easiest to investigate is the invariant sector: $A^+_1$.
 We therefore focus on this particular symmetry sector.
 In the first order approximation,
 if we neglect the mixing between
 the $s$-wave and $g$-wave, we have for the $A^+_1$ sector:
 \be
 \label{eq:swave_simple}
 \cos(\Delta_1+\Delta_2-\delta^0_1-\delta^0_2)=
 \eta_0\cos(\Delta_1-\Delta_2-\delta^0_1+\delta^0_2)\;.
 \ee
 where we have also used the special
 parametrization~(\ref{eq:S_parametrize})
 for the $s$-wave $S$-matrix elements.
 The quantities appearing above,
 namely  $\delta^0_1$,  $\delta^0_2$, $\eta_0$,
 $\Delta_1$ and $\Delta_2$, are all functions of the energy:
 $E=k^2_1/(2m_1)=E_T+k^2_2/(2m_2)$.
 Eq.~(\ref{eq:swave_simple}) is the simplified formula for the $s$-wave
 $S$-matrix elements when contaminations
 from higher angular momentum (mainly from $l=4$) are neglected.
 This relation is helpful since
 it provides a constraint on the four physical quantities.
 If the mixing of the $g$-wave is taken into account, the
 corresponding formula becomes more complicated~\cite{chuan05:2channel}.
 Although we worked out the formulae in a cubic box, similar relations can also
 be obtained for general rectangular box following
 the strategies outlined in
 Ref.~\cite{chuan04:asymmetric}.

 Finally, let us speculate about possible extension to
 the case of massive field theory. As in the case of single
 channel scattering~\cite{luscher91:finitea,kim05:moving}, one can work
 in the Bethe-Salpeter formalism for the two particle states
 and the results reported here can possibly be generalized
 to field theory. Details for this are now under investigation.

 \section{Conclusions}
 \label{sec:conclude}

 In this work, we have studied two-particle two-channel
 scattering states in a cubic box with periodic boundary conditions.
 Assuming that energy eigenstates are only two-particle states,
 the relation of the exact energy eigenvalues in the box and the
 physical parameters in the coupled channel $S$-matrix elements
 in the continuum is found. This formula
 can be viewed as a generalization of
 the well-known L\"uscher's formula to the
 coupled channel situation (inelastic scattering).
 In particular, we show that the two-channel $S$-matrix elements
 in the $s$-wave are related to the energy of the two-particle system
 by a simple identity, if contaminations from higher
 angular momentum sectors are neglected.
 This relation is non-perturbative in nature and it
 will help us to establish connections between the
 $S$-matrix parameters in the multi-channel
 scattering with the energy eigenvalues which are
 in principle accessible in lattice calculations.

 \section*{Acknowledgments}

  We would like to thank H.~Q.~Zheng for helpful discussions.


\begin{thebibliography}{0}
\bibitem{E86500:pipi}
P.~Tru{\"o}l,
\newblock {\em hep-ex/0012012}, 2000;
M.J. Matison et~al.,
\newblock {\it Phys. Rev. D} {\bf 9}, 1872 (1974);
N.O. Johannesson and J.L. Petersen,
\newblock {\it Nucl. Phys. B} {\bf 68}, 397 (1973);
A.~Karabouraris and G.~Shaw,
\newblock {\it J. Phys. G} {\bf 6} 583 (1980);
A.D. Martin,
\newblock {\it Nucl. Phys. B} {\bf 179}, 33 (1981).

\bibitem{luscher91:finitea}
M.~L{\"u}scher,
\newblock {\it Commun. Math. Phys.} {\bf 105}, 153 (1986);
M.~L{\"u}scher and U.~Wolff,
\newblock {\it Nucl. Phys. B} {\bf 339}, 222 (1990);
M.~L{\"u}scher,
\newblock {\it Nucl. Phys. B} {\bf 354}, 531 (1991);
M.~L{\"u}scher,
\newblock {\it Nucl. Phys. B} {\bf 364}, 237 (1991).

\bibitem{Zimmermann94}
M.~Goeckeler, H.A. Kastrup, J.~Westphalen, and F.~Zimmermann,
\newblock {\it Nucl. Phys. B} {\bf 425}, 413 (1994).

\bibitem{gupta93:scat}
R.~Gupta, A.~Patel, and S.~Sharpe,
\newblock {\it Phys. Rev. D} {\bf 48}, 388 (1993);
M.~Fukugita, Y.~Kuramashi, H.~Mino, M.~Okawa, and A.~Ukawa,
\newblock {\it Phys. Rev. D} {\bf 52}, 3003 (1995);
S.~Aoki et~al.,
\newblock {\it Nucl. Phys. (Proc. Suppl.) B} {\bf 83}, 241 (2000);
S.~Aoki et~al.,
\newblock {\it Phys. Rev. D} {\bf 66}, 077501 (2002);
C.~Liu, J.~Zhang, Y.~Chen, and J.P. Ma,
\newblock {\it Nucl. Phys. B} {\bf 624}, 360 (2002);
S.~Aoki et~al.,
\newblock {\it Phys. Rev. D} {\bf 67} 014502 (2003).

\bibitem{CPPACS03:pipi_phase_unquench}
T.~Yamazaki et~al.,
\newblock {\it Phys. Rev. D} {\bf 70} 074513 (2004).

\bibitem{chuan05:2channel} S. He, X. Feng and C. Liu {\it JHEP}
{\bf 0507}, 011 (2005).

\bibitem{chuan04:asymmetric}
X.~Li and C.~Liu,
\newblock {\it Phys. Lett. B} {\bf 587} 100 (2004);
X.~Feng, X.~Li, and C.~Liu,
\newblock {\it Phys. Rev. D} {\bf 70} 014505 (2004).
\bibitem{kim05:moving}
C.h. Kim, C.T. Sachrajda and S. Sharpe,
\newblock {\it hep-lat/0507006} 2005.

\end{thebibliography}
\end{document}